\documentclass[a4paper,twocolumn,pra,reprint,longbibliography,superscriptaddress,nofootinbib,hyphens]{revtex4-2}

\usepackage{amssymb,amsmath,bbm}
\usepackage{mathtools,hyperref}
\usepackage{physics,braket}
\usepackage{graphicx,xcolor}
\usepackage[varg]{txfonts}

\newcommand{\sfrac}[2]{\text{\small $\frac{#1}{#2}$}}
\newcommand{\cm}{\text{\tiny CM}}

\begin{document}

\title{Searching for a physical description relative to a quantum system}

\author{Henrique A. R. Knopki\href{https://orcid.org/0009-0001-2260-8784}{\includegraphics[scale=0.05]{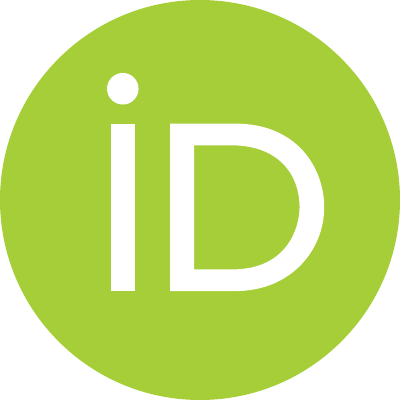}}}
\email{hknopki@alumni.usp.br}

\author{Renato M. Angelo\href{https://orcid.org/0000-0002-7832-9821}{\includegraphics[scale=0.05]{orcidid.pdf}}}
\email{renato.angelo@ufpr.br}

\affiliation{Department of Physics, Federal University of Paraná, P. O. Box 19044, Curitiba, 81531-980, Paraná, Brazil}
            
\begin{abstract}
Physics is a model of nature able to both describe and predict the results of measurements made with respect to reference systems. These reference systems, in turn, are themselves physical and thus subject to the laws of physics. The situation is no different when the model in use is quantum mechanics: states and observables are relative entities, and reference frames are not exempt from exhibiting quantum behavior. In recent years, the scientific community has shown renewed interest in quantum reference frames, particularly in connection with the covariance of physical laws and quantum resources. However, current approaches fall short of providing a complete prescription for predicting observables associated solely with degrees of freedom accessible from the quantum reference frame. In pursuit of such a description, we show that while this is fully feasible for two-particle systems, there are irreducible difficulties that arise in many-body systems. In particular, within the framework of Galilean relativity, with absolute time, we demonstrate that a canonical and relational description with respect to a particle in the system cannot be achieved through any unitary transformation. Our findings call for new strategies to address the problem of quantum reference frames.
\end{abstract}

\maketitle

\section{Introduction} 
\label{sec:intro}

Empirical observation is always performed with respect to some, often idealized, physical system---a reference frame. And, according to the principle of relativity, there must be no privileged inertial reference frames; i.e., the fundamental physical laws are perspective-independent. The question then naturally arises as to whether quantum systems can play the role of reference frames and how physics looks from their perspective. A seminal step toward answering this question was given by Aharonov and Kaufherr~\cite{Kaufherr1984}, who showed that quantum mechanics can be coherently formulated from a particle's perspective---a quantum reference frame (QRF). The philosophy of their work was to fully embed the theory of reference frames into quantum theory, eliminating the necessity for a classical reference frame.

The relevance of QRFs has been recognized in a variety of contexts, ranging from fundamental aspects of physics~\cite{Aharonov1967,Kitaev2004,Bartlett2006,Angelo2011,Angelo2012,Angelo2015,Loveridge2017,Giacomini2019}, passing through quantum information \cite{Spekkens2007,Bartlett2014,Angelo2021,Cepollaro2024}, quantum communication~\cite{Popescu1995,Bartlett2003, Turner2009}, and entanglement detection~\cite{Brukner2009,Rudolph2010}, up until resource theories~\cite{Spekkens2008} and thermodynamics~\cite{Winter2018}. The role of QRFs has also been extended to contexts including relativistic quantum theory \cite{Terno2002,Adami2002,Terno2004,Brukner2019}, quantum gravity \cite{Rovelli1991,Rovelli1991_2,Dittrich2006,Hohn2019,Ruiz2020,Brukner2020}, and quantum field theory \cite{Mann2020,Brukner2020_2}.

Today, the literature on QRFs has reached a high level of maturity, with numerous conceptual and formal developments~\cite{Giacomini2019,Ruiz2020,Merriam2005,Hamette2020,Krumm2021,Hoehn2022,Hoehn2019,Vanrietvelde2023,Hoehn2020b,Hoehn2021a,Giacomini2019b,Streiter2021,Ballesteros2021,Giacomini2021,Smith2020,Mikusch2021,Hoehn2022b,CastroRuiz2020,Hoehn2021b,Lake2023,Castro2025,Carette2025}. Some of these approaches are based on the premise that a change in perspective is enabled by unitary transformations, and that different degrees of freedom may necessitate distinct transformations. This idea is quite defensible in ideal scenarios because the transformation between reference frames is a purely theoretical maneuver; we should always be able to recover the original description reversibly. Nevertheless, as we will argue in this work, certain subtleties persist in the unitary transformations known to date, which prevent achieving a truly relative description from the perspective of a QRF for a multiparticle system. This result calls for a careful review of certain interpretative and operational aspects established by the approaches available up to now.

The assertions above will be substantiated through the following structure. In Sec.~\ref{sec:section2}, we will discuss in detail what we understand by the transformation between quantum reference frames, highlighting the differences from other approaches. Subsequently, in Sec.~\ref{sec:some_Ts} we provide a brief review of two well-established transformations, namely, one that maps laboratory coordinates to the center-of-mass and relative coordinates and another that maps to relative coordinates only. Then, by applying both transformations to a concrete problem in Sec.~\ref{sec:paradox}, we will encounter a paradox that serves precisely to highlight the subtle issue we mentioned above. In Sec.~\ref{sec:notransf}, by resolving the paradox, we present our main result, namely, that there is no transformation capable of achieving a genuinely relational transformation (in the sense that we will define here). Section~\ref{sec:discussion} is then dedicated to our conclusions.

\section{Reference frame transformations} 
\label{sec:section2}

Physical events are reference-frame-independent phenomena; their locations in space-time, however, are not. For instance, a detector click is a fact that occurs for every observer, though at different space-time points. Likewise, although the probability of a detector click occurring is invariant under reference frame transformations, the probability of it occurring at a specific space-time location is not. However simple this remark may seem, it merits elaboration. Considering $T$ as the unitary representation of a coordinate transformation from a reference frame $S$ to another frame $S'$, it is not uncommon to find in the literature the following prescription for the expectation value of a generic observable $O$: 
\begin{equation}\label{eq:<O>=<O>'}
\bra{\psi}O\ket{\psi} = \bra{\psi}T^{\dagger}TOT^{\dagger}T\ket{\psi} = \bra{\psi'}O'\ket{\psi'},
\end{equation}
where $O' = T O T^{\dagger}$ and $\ket{\psi'} = T \ket{\psi}$ represent how the frame $S'$ describes the observable and the state vector in terms of the physical quantities accessible to it. If $O$ is a measurement operator, then relation \eqref{eq:<O>=<O>'} implies the frame invariance of probability distributions. An immediate consequence of this formulation is that Bell inequalities---such as the Clauser-Horne-Shimony-Holt (CHSH) inequality~\cite{chsh}---which are based on expectation values of observables, remain invariant under reference frame transformations, including Lorentz transformations~\cite{BellFlaminia,BellInvariant}.

Although the interpretation of $O'$ and $\ket{\psi'}$ is standard and widely accepted, the meaning of $\braket{\psi'|O'|\psi'}$ warrants clarification. To see why, consider the transformation $T_G = e^{i (m v X - v t P) / \hbar}$, where $X$ is the position operator of a particle of mass $m$, $P$ is the momentum operator, and $v$ and $t$ are scalars with units of velocity and time, respectively. Let us further assume that $\ket{\psi}$ is a Gaussian state centered at position $x$. While the transformed position operator reads $X'=X-vt \mathbbm{1}$, we find that $\braket{\psi'|X'|\psi'} = x$, in contrast to the result $x - v t$ expected under a Galilean boost. Therefore, $\braket{\psi'|O'|\psi'}$ cannot be regarded as the expectation value of a measurement performed in $S'$. Rather, it is merely a reformulation of the expectation value computed in $S$, expressed in terms of the `primed objects' accessible from $S'$.

In this work, we adopt the following formulation:
\begin{equation}\label{eq:<O>'}
\expval{O}'=\bra{\psi}T^{\dagger} O T\ket{\psi},
\end{equation}
where $\expval{O}'$ is to be interpreted as the expectation value measured in $S'$. Indeed, specializing $T \equiv T_G^\dagger$ in the Galilean boost case yields $\expval{X}' = \expval{X} - v t$, which relates the expectation value $\expval{X}'$ measured in $S'$ to the expectation value $\expval{X} = \expval{X}{\psi}$ measured in $S$. As pointed out in Refs.~\cite{Angelo2015,Angelo2021}, the description \eqref{eq:<O>'} admits alternative frameworks: an active picture, $\expval{O}'=\expval{O}{\psi'}$, in which the state is transformed but the observables are not, and a passive picture, $\expval{O}'=\expval{O'}{\psi}$, in which the opposite occurs. It should be clear from this latter formula that the prescription in \eqref{eq:<O>'} refers to an experiment involving measurements of $O'$, a physical quantity accessible from $S'$. Thus, $\expval{O}'$ is not merely a theoretical rephrasing of the experiment conducted in $S$. This is why this type of formulation predicts that violations of Bell inequalities are not invariant under Lorentz transformations~\cite{Matsas,ThreeParticleBellNotInvariant,BellNotInvariant}: this description refers to an experiment in which the detector velocity changes according to the boost applied.

\section{Some meaningful unitary transformations} 
\label{sec:some_Ts}
In this section, we review some well-known frame transformations, examining both the active and passive pictures, with the aim of gaining physical intuition.

\subsection{Center-of-mass and relative coordinates}
\label{sec:T_cm,r}

Let us consider a two-particle system consisting of particles with masses $m_0$ and $m_1$, described in $S$ (an inertial laboratory frame) by one-dimensional coordinates $x_0$ and $x_1$. The quantum mechanical counterparts of these coordinates, together with their conjugate momenta $P_0$ and $P_1$, are the position operators $X_0$ and $X_1$. It is straightforward to verify that the transformation $T_{\cm,r}:\mathcal{H}_0\otimes\mathcal{H}_1 \to \mathcal{H}_0\otimes\mathcal{H}_1$, explicitly defined by the unitary operator~\cite{Angelo2012}
\begin{equation} \label{eq:T_cm,r}
T_{\cm,r}=\exp\left(-\sfrac{i}{\hbar}\sfrac{m_1}{M}X_1P_0\right)\exp\left(\sfrac{i}{\hbar}X_0P_1\right),
\end{equation}
implements the change of description from the laboratory frame $S$ to the center-of-mass (CM) and relative ($r$) coordinates. (Throughout this article, expressions like $X_1P_0$ are shorthand for $X_1 \otimes P_0$, and similar notation applies to other terms.) In the passive picture, we use the notation $O_{0,1}' \equiv T_{\cm,r}^\dag O_{0,1} T_{\cm,r}$, from which one obtains
\begin{subequations}\label{eq:01->cmr}
\begin{align}
    &X_0'=\sfrac{m_0 X_0 + m_1 X_1}{M}, 
    &&P_0' = P_0 + P_1,  \\
    &X_1' = X_1 - X_0, 
    &&P_1' = \mu_{01} \qty( \sfrac{P_1}{m_1} - \sfrac{P_0}{m_0}),
\end{align}
\end{subequations}
where $M = m_0 + m_1$ and $\mu_{01} = m_0 m_1 / M$. One can thus make the obvious identifications $X_0' \equiv X_\cm$, $P_0' \equiv P_\cm$, $X_1' \equiv X_r$, and $P_1' \equiv P_r$, which are the well-known results for the two-particle problem. In the active picture, we find that
$T_{\cm,r}\ket{a}_0\ket{b}_1 = \ket{\tfrac{m_0 a + m_1 b}{m_0 + m_1}}_0 \ket{b - a}_1$, showing that the position eigenstates $\ket{a}_0$ and $\ket{b}_1$ in frame $S$ are mapped to kets that encode the center-of-mass and relative coordinates, respectively. Since $T_{\cm,r}$ does not map the vector out of its Hilbert space, we might omit the subscripts $0$ and $1$ in the vectors. However, it is convenient to indicate the interpretation we assign to each ket state. We therefore employ the notation
\begin{equation} \label{eq:ket01->ketcmr}
T_{\cm,r}\ket{a}_0\ket{b}_1=\left|\tfrac{m_0a+m_1b}{m_0+m_1}\right\rangle_\cm\ket{b-a}_r,
\end{equation}
which assumes the relabeling $(0,1)\to (\text{CM},r)$. The application of the above prescription to a generic state $\ket{\psi} = \int\mathrm{d}u\mathrm{d}v\,\psi(u,v) \ket{u}_0\ket{v}_1$ gives
\begin{equation}\label{eq:psi'_cm,r}
\ket{\psi'}_{\cm,r} = T_{\cm,r}\ket{\psi}=\int\mathrm{d}u\mathrm{d}v\,\psi_{\cm,r}'(u,v) \ket{u}_\cm\ket{v}_r,
\end{equation}
in which $\psi_{\cm,r}'(u,v)=\psi\qty(u-\frac{m_1}{M}v, u+\frac{m_0}{M}v)$ represents the transformed wave function. 

It is worth noting that the new coordinate system is, so to speak, a hybrid one: the CM position remains a physical variable relative to the laboratory frame $S$, whereas $x_r$ corresponds to the position of particle 1 relative to particle 0---the latter thus playing the role of the quantum reference frame $S'$.

\subsection{The relational approach} 
\label{sec:T_R}

Introduced in Ref.\cite{Giacomini2019}, and in a somewhat equivalent form in Ref.\cite{Kaufherr1984}, the relational ($R$) approach aims to describe physics in a way that equalizes the number of degrees of freedom accessible from each quantum reference frame. In this framework, if $S$ describes the physics of particles $0$ and $1$, then particle $0$, when promoted to the role of reference frame $S'$, is allowed to describe the physics of $S$ and $1$. However commendable this approach may be, the authors of Ref.~\cite{Giacomini2019} adopted a conceptual attitude consistent with that underlying Eq.~\eqref{eq:<O>=<O>'}. We, on the other hand, pursue a framework as described by Eq.~\eqref{eq:<O>'}. In our prescription, the transformation $T_R: \mathcal{H}_0 \otimes \mathcal{H}_1 \rightarrow \mathcal{H}_0 \otimes \mathcal{H}_1$, which maps the spatial descriptions of particles $0$ and $1$ relative to $S$ into spatial descriptions of the laboratory $S$ and particle $1$ relative to particle $0$, is given by
\begin{equation} \label{eq:T_R}
T_R=\pi_0\exp(\sfrac{i}{\hbar}P_1X_0).
\end{equation}
The parity operator $\pi_0:\mathcal{H}_0\mapsto \mathcal{H}_0$ acts according to $\pi^{\dagger}_0(X_0,P_0)\pi_0=(-X_0,-P_0)$. In the context of the passive picture, we adopt the notation $O_{0,1}' \equiv T_R^\dag O_{0,1} T_R$, from which one obtains
\begin{subequations}\label{eq:01->R}
\begin{align}
& X_0' =-X_0, 
&& P_0' = -(P_0 + P_1),\\
& X_1' = X_1-X_0,
&& P_1' = P_1.
\end{align}
\end{subequations}
Although we correctly obtained the correspondences $X_0'\equiv X_S$ (position of $S$ relative to particle $0$) and $X_1'\equiv X_r$  (position of particle $1$ relative to particle $0$), it is evident that the momenta $P_{0,1}'$ do not correspond to objects relative to particle $0$. In fact, $P_0'$ is the negative of the center-of-mass momentum, whereas $P_1'$ is nothing but the momentum of particle $1$ relative to $S$. In the active picture, a direct calculation shows that $T_R\ket{a}_0\ket{b}_1 = \ket{-a}_0\ket{b-a}_1$, which, using the notation adopted in this work, can be written as 
\begin{equation}\label{eq:ket01->ketR} 
T_R\ket{a}_0\ket{b}_1 = \ket{-a}_S\ket{b-a}_r. 
\end{equation}
The application of the relational prescription to a generic state $\ket{\psi} = \int\mathrm{d}u\mathrm{d}v\,\psi(u,v) \ket{u}_0\ket{v}_1$ gives
\begin{equation} \label{eq:psi'_R}
    \ket{\psi}_R' = T_R \ket{\psi}
    = \int\mathrm{d}u\mathrm{d}v\,\psi'_R(u,v) \ket{u}_S\ket{v}_r,
\end{equation}
where $\psi_R'(u,v) = \psi(-u,v-u)$ represents the relational wave function. In Ref.~\cite{Angelo2021}, for the instance in which $S$ describes the physics of a single particle $0$, the authors propose to take $T_R = \pi_0$ as the relational transformation. Clearly, this implements the desired spatial relationality, for if $\ket{a}_0$ gives the location $a$ for particle $0$ relative to $S$, then $\pi_0\ket{a}_0 =\ket{-a}_S $ gives the location $-a$ for $S$ relative to particle $0$.

\section{Paradox}
\label{sec:paradox}

\begin{figure}[t]
    \centering
    \includegraphics[scale=0.37]{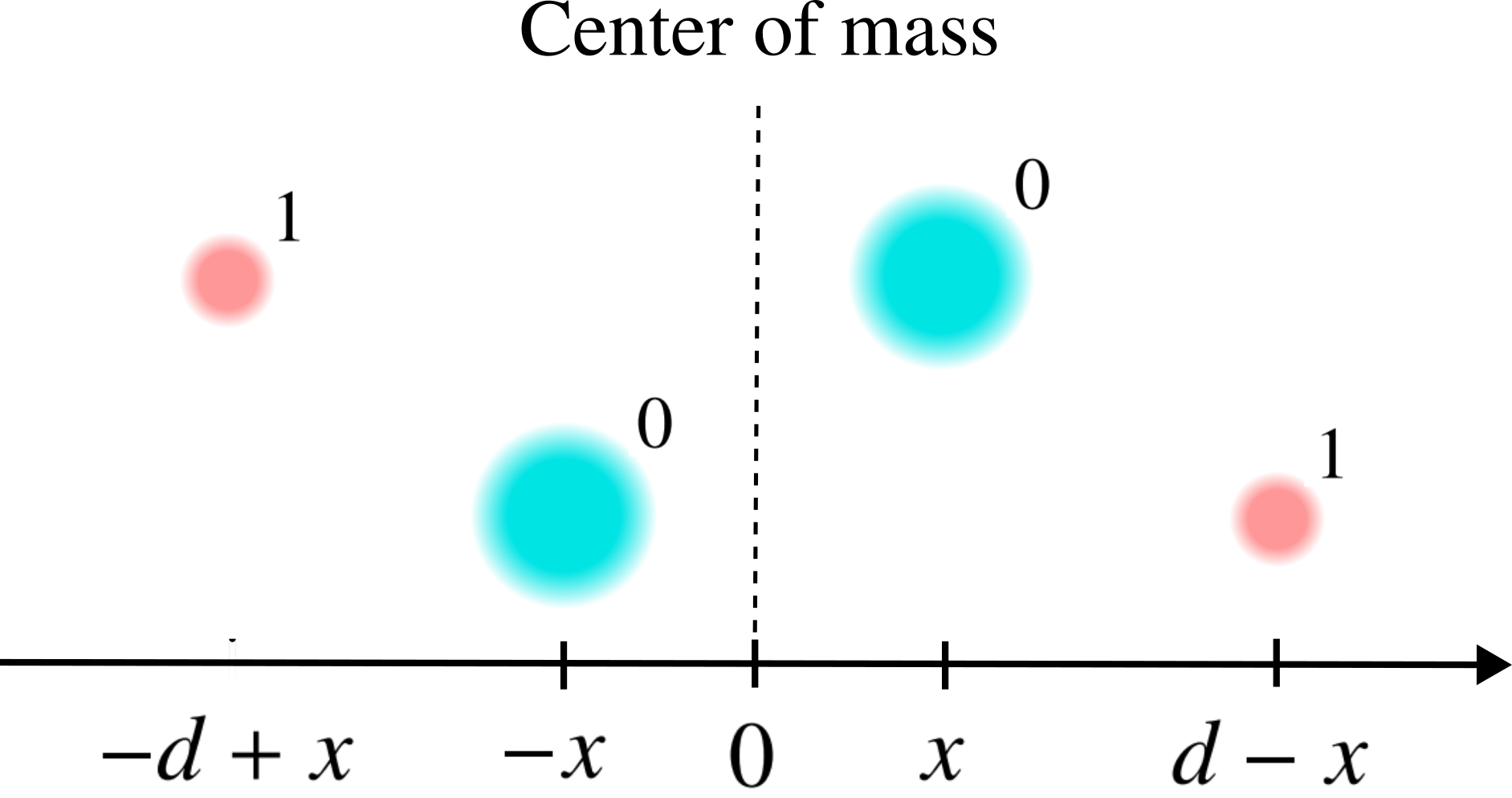}
    \caption{Classically, the decay can occur with particle $0$ occupying position $x$ and particle $1$ position $-d + x$ (top line), or with particle $0$ occupying position $-x$ and particle $1$ position $d - x$ (bottom line). Quantum mechanically, the two scenarios can exist in superposition [see Eq.~\eqref{eq:statedecay}]. As the system is subject solely to internal forces, its center of mass must remain fixed at the origin of the coordinate system, implying that $x = d m_1 / M$.}
    \label{fig:1}
\end{figure}

Both transformations considered thus far correctly yield the relative coordinate $X_r = X_1 - X_0$, suggesting that either may be valid when only this relative coordinate is of interest. To test this intuition, we consider a physical system in which a parent particle decays into daughter particles $0$ and $1$ in a quantum superposition, as illustrated in Fig.~\ref{fig:1}. In the laboratory frame $S$, we have:
\begin{equation} \label{eq:statedecay}
    \ket{\psi}=\frac{\ket{x}_0 \ket{-d+x}_1 + e^{i\phi}\ket{-x}_0 \ket{d-x}_1}{\sqrt{2}},
\end{equation}
where to each ket $\ket{\bar{z}}_{n}$ above is associated a Gaussian spatial amplitude $\sqrt{G_{\bar{z},\sigma_n}(z)}$, with $n\in\{0,1\}$, $z\in\{u,v\}$, and $G_{\bar{z},\sigma_n}(z)=\exp[-(z-\bar{z})^2/(2\sigma_n^2)]/\sqrt{2\pi\sigma_n^2}$, whose spatial uncertainty $\sigma_n$ is much smaller than the separation between the corresponding packets, and $\phi$ is an arbitrary phase. Since the center of mass remains at rest upon the decay, $x = d m_1 / M$.
The application of the reference frame transformations $T_{\cm,r}$ and $T_R$ leads to the following forms of the state vector:
\begin{subequations}\label{eq:decay}
\begin{align}
\ket{\psi}_{\cm,r}&=\ket{0}_\cm\qty(\frac{\ket{-d}_r+e^{i\phi}\ket{d}_r}{\sqrt{2}}), \label{eq:decay_cm,r}\\
\ket{\psi}_R&=\frac{\ket{-x}_S\ket{-d}_r+e^{i\phi}\ket{x}_S\ket{d}_r}{\sqrt{2}}. \label{eq:decay_R}
\end{align}
\end{subequations}
At first sight, the two descriptions appear to agree: measurements of the relative position of particle 1 yield outcomes centered around $-d$ and $d$, with equal probability in both treatments. Also, $\braket{\psi|X_1|\psi}_{\cm,r} = \braket{\psi|X_1|\psi}_R = 0$. However, momentum measurements reveal the relative phase $\phi$ in only one of the two scenarios. For instance, consider the shift operator $e^{-i(2d)P_1/\hbar}$, which is commonly used in discussions of modular momentum~\cite{Aharonov2005}. Measurements of this operator (or of its real part) would yield the expectation values
\begin{subequations}
\begin{align}
&\Tr\left(e^{-i(2d)P_1/\hbar}\ket{\psi}\!\bra{\psi}_{\cm,r}\right)=e^{i\phi}/2, \\
&\Tr\left(e^{-i(2d)P_1/\hbar}\ket{\psi}\!\bra{\psi}_R\right)=0.
\end{align}
\end{subequations}
The capacity of encoding the phase $\phi$ is already apparent in the reduced state associated to the relative coordinate. By tracing out the $\text{CM}$ and $S$ degrees of freedom, we obtain
\begin{subequations} \label{eq:rho_r}
\begin{align}
\rho_r^{(\cm,r)}&=\tfrac{1}{2}\Big(\ket{-d}_r+e^{i\phi}\ket{d}_r\Big)\Big(\bra{-d}_r+e^{-i\phi}\bra{d}_r\Big),\\
\rho_r^{(R)}&=\tfrac{1}{2}\Big(\ket{-d}\!\bra{-d}_r+\ket{d}\!\bra{d}_r\Big).
\end{align}
\end{subequations}
Clearly, the incoherent mixture emerging in the relational approach cannot encode the relative phase $\phi$, which is a consequence of the fact that the relative coordinate is entangled with the position of the laboratory $S$ relative to particle $0$. Here is the paradox: after applying two valid canonical transformations, both involving the relative coordinate, we arrive at the descriptions \eqref{eq:decay_cm,r} and \eqref{eq:decay_R}, which yield distinct predictions for the physics relative to the quantum reference frame $S'$ (particle $0$).

Of course, there is nothing mathematically incorrect about either $T_{\cm,r}$ or $T_R$. From a theoretical standpoint, one is free to choose any set of coordinates. The issue lies in identifying which one (if any) faithfully encodes the relative physics. To this end, it is key to recall that when writing states such as those in Eqs.~\eqref{eq:rho_r}, we are not merely representing positional information, but rather algebraic objects that also encapsulate information about the corresponding conjugate momenta. It then turns out that, while $\rho_r^{(\cm,r)}$ refers to the conjugate pair $\big\{X_1', P_1'\big\} = \Big\{X_1 - X_0, \mu_{01}\Big(\frac{P_1}{m_1} - \frac{P_0}{m_0}\Big)\Big\}$, which clearly captures the physics of genuinely relative position and momentum variables, $\rho_r^{(R)}$, instead, encodes information about the exotic canonical pair $\big\{X_1', P_1'\big\} = \big\{X_1 - X_0, P_1\big\}$, a pair that relates the relative position to the original momentum. In other words, the physics captured by $\rho_r^{(R)}$ is ``hybrid" within its very canonical structure: while the position is expressed relative to $S'$, the momentum is defined relative to $S$.

Another fundamental issue underlies the relational approach. In deriving Eqs.~\eqref{eq:decay}, we have tacitly assumed that the relation $T_R\ket{a}_0\ket{b}_1 = \ket{-a}_S\ket{b-a}_r$, which clearly applies to position eigenstates, also holds for Gaussian states. However, the transformation of $\psi(u,v)=\sqrt{G_{a,\sigma_0}(u)}\sqrt{G_{b,\sigma_1}(v)}$ directly yields $\psi_R'(u,v)=\psi(-u,v-u)=\sqrt{G_{-a,\sigma_S}(u)}\sqrt{G_{b-a,\sigma_1}(v)}\,\Gamma(u,v)$, where
\begin{align}
 \Gamma(u,v)&=\exp\qty[\sfrac{(u+a)(v-b+a)}{2\sigma_1^2}],
\end{align}
and $\sigma_S=\sigma_0\sigma_1/\qty(\sigma_0^2+\sigma_1^2)^{1/2}$. In contrast with the transformation involving center-of-mass and relative coordinates~\cite{Angelo2011}, the relational approach does not admit any reasonable regime in which the crossing term $\Gamma(u,v)$ can be neglected. Therefore, we should rather have used the notation $T_R\ket{a}_0\ket{b}_1 = \ket{-a, b-a}_{S,r}$ to stress that physical states acquire ``intrinsic entanglement'' upon the active transformation. Of course, this is not just a notational issue: in Eq.~\eqref{eq:decay_R}, e.g., this means that entanglement would persist even when $x=d\qty(\frac{m_1}{m_1+m_0}) \to 0$, which occurs when $\frac{m_1}{m_0} \to 0$. In this case, however, the CM and particle $0$ effectively become the same system, whose position could not get entangled with the relative coordinate through internal interactions.


One may wonder whether the transformation $T_{\cm,r}$, which has been shown to successfully furnish the relative sector for two-particle systems, admits a generalization to an $N$-particle system. A natural attempt is given by
\begin{align}
    T_{\cm,r}^{(N)} = \exp(-\sfrac{i}{\hbar}\sum^{N-1}_{i=1}\sfrac{m_i}{M}X_iP_0)\exp(\sfrac{i}{\hbar} \sum^{N-1}_{i=1}X_0P_i).
\end{align}
However, by applying this transformation we find
\begin{subequations}
\begin{align}
& X_0'=\sum^{N-1}_{k=0} \sfrac{m_k}{M}X_k, 
&& P_0'= \sum^{N-1}_{k=0} P_k, \\
& X_i' = X_i-X_0, 
&& P_i' = m_i\qty(\sfrac{P_i}{m_i}-\sfrac{P_\cm}{M}),
\end{align}
\end{subequations} 
where the integer $i$ is restricted, by our convention, to the domain $\{1,2,\cdots,N-1\}$. Although the results allow for the desired identifications $X_0'\equiv X_\cm$, $P_0'\equiv P_\cm$, and $X_i'\equiv X_{r_i}$, we see that $P_i'\neq \mu_{i0} \qty(\frac{P_i}{m_i} - \frac{P_0}{m_0})\equiv P_{r_i}$, where $\mu_{i0} = m_i m_0 / (m_i + m_0)$. That is, the momenta $P_i'$ conjugate to $X_{r_i}$ are defined relative to the CM, not to particle $0$. Again, this represents a hybrid canonical structure: the positions $X_i'$ are defined relative to $S'$, while the momenta $P_i'$ are defined relative to the CM.

It is well known~\cite{Angelo2012,Angelo2015} that there exists a transformation that yields the momenta $P_{r_i}$ relative to $S'$; however, the corresponding conjugate coordinates are not relative to that frame.

\section{No transformation to a full relative description} 
\label{sec:notransf}

We now demonstrate, by \textit{reductio ad absurdum}, that no transformation  can provide a canonical description that is exclusively relative to $S'$ for an $N$-particle system. Let us begin by assuming that there exists a unitary transformation $T$ that yields the desired relations
\begin{align}\label{eq:relativevariables}
    &X_i'=X_i-X_0,
    &P_i'=\mu_{i0}\qty(\sfrac{P_i}{m_i}-\sfrac{P_0}{m_0}).
\end{align}
We make no \textit{a priori} assumption about the transformed variables $X_0' = T^\dag X_0 T$ and $P_0' = T^\dag P_0 T$, which may correspond to center-of-mass physics relative to $S$, to the laboratory’s degrees of freedom relative to $S'$, or possibly to something else entirely. Since $T$ is unitary by hypothesis, we have $[X_i', P_j'] = T^\dag [X_i, P_j] T = i\hbar \delta_{ij}T^\dag T = i\hbar \delta_{ij}\mathbbm{1}$, for any integer $0\leq j\leq N-1$. When $j\geq 1$, it follows from the relations~\eqref{eq:relativevariables} that
\begin{align}\label{eq:absurdum}
    i\hbar\delta_{ij}\mathbbm{1}&=[X_i-X_0,\mu_{j0}\qty(\sfrac{P_j}{m_j}-\sfrac{P_0}{m_0})] \nonumber \\
    &= \sfrac{m_0}{m_j+m_0}i\hbar\delta_{ij}\mathbbm{1}+\sfrac{m_j}{m_j+m_0}i\hbar\mathbbm{1} \nonumber \\
    &= i\hbar\mathbbm{1}\qty(\sfrac{m_j+m_0 \delta_{ij}}{m_j+m_0}).
\end{align}
Upon further manipulation, the above equality gives $\delta_{ij} = 1$, which leads to an absurdity when $i \neq j$. Therefore, a unitary transformation $T$ that satisfies the relations (\ref{eq:relativevariables}) cannot exist.
 
Note that the \textit{absurdum} in Eq.~\eqref{eq:absurdum} disappears in the limit $\frac{m_j}{m_0} \to 0$. In this regime, particle $0$ experiences no appreciable back-reaction from measurements performed on the other particles, so that the aspects of measurement incompatibility encoded in the commutation relation resume their usual form, $[X_i', P_j'] = i\hbar \delta_{ij} \mathbbm{1}$, typical of infinite-mass inertial reference frames.

Recently, in an effort to derive reference-frame transformations for subsystems, a framework has been put forward~\cite{Castro2025} for $N$-particle systems wherein particles $0$ and $1$ are taken as references for position and velocity, respectively, of the remaining $N-2$ particles. In this case, the relative physics is described in terms of canonically conjugate pairs of relative position and momentum operators, $\qty(X_{k|1}', P_{k|2}')$, where $X_{k|1}' = X_k - X_1$ and $P_{k|2}' = P_k - \frac{m_k}{m_2} P_2$, for $k \geq 3$. However useful and promising it may be, this approach does not solve the problem under inspection here, as it remains hybrid in the sense discussed above.

\section{Discussion}
\label{sec:discussion}

We have thus far demonstrated that no unitary transformation can map the original set of variables $\mathbf{R}=\{X_0, P_0, X_1, P_1,\cdots,X_{N-1},P_{N-1}\}$ to the set $\mathbf{R}'=\{-X_0, -P_0, X_{r_1},P_{r_1}, \cdots,X_{r_{N-1}},P_{r_{N-1}}\}$, which would correspond to a description in which particle $0$ serves as the QRF. This constitutes an important no-go result, challenging the widespread belief that a covariant relational description could be obtained through unitary transformations.

It should be emphasized, however, that this does not imply that the concept of QRF is fundamentally flawed, or that the predictions made from such frames are inscrutable. In fact, the idea of a QRF---especially within the scope of Galilean relativity---is entirely grounded in the premise that spatial coordinates depend on kinematic relationality, regardless of whether the mechanics under consideration is quantum or classical. The entire literature on QRFs relies on this postulate, which demands a critical review of some classical dogmas. For example, if a particle $0$ is prepared, relative to the laboratory frame $S$, in a superposition $\ket{\psi} = \int\mathrm{d}u\,\qty[G_\sigma(u - d) + G_\sigma(u + d)]\ket{u}_0$ of Gaussian packets $G_{\pm d,\sigma}(u)$ centered at $\pm d$ with width $\sigma$, then there is no theoretical reason to exclude the particle from describing $S$ as being in the superposition $\ket{\psi'} = \pi_0\ket{\psi} = \int\mathrm{d}u\,\qty[G_\sigma(u + d) + G_\sigma(u - d)]\ket{u}_S$. In particular, the measurement results for the particle $0$ relative to $S$ are also the measurement results of the frame $S$ relative to the particle $0$. Therefore, in this context, the classical notion of absolute spatial localization for macroscopic systems can no longer be upheld. Once we accept that this is how nature operates, it becomes meaningful to ask how quantum systems describe the world.

We began by arguing that $\braket{O}' = \braket{\psi | T^\dag O T | \psi}$ is the appropriate prescription for grasping the physics from the perspective of a QRF. Although this is strictly correct for a two-particle system, we have shown that this structure cannot be maintained in general, since no unitary transformation can be found that assigns the role of QRF to one particle in many-particle systems. In any case, by the consistency of the arguments presented, we postulate that the appropriate prescription can be maintained as $\braket{f}' = \braket{\psi | f (\mathbf{R}') | \psi}$; or, for more general states,
\begin{equation}\label{eq:prescription}
    \braket{f}'=\Tr\qty[f(\mathbf{R}')\,\rho],
\end{equation}
for an arbitrary (well behaved) function $f$. As an example of how this recipe can be applied, consider the symmetrized action function $f(\mathbf{R}') = \frac{1}{2}\qty(X_1' P_2' + P_2' X_1')$, which may be interpreted as a component of orbital angular momentum, provided that degree of freedom $2$ corresponds to a distinct spatial direction of the same particle. By using the commutation relation \eqref{eq:absurdum}, one finds $f(\mathbf{R}') = X_1' P_2' - \frac{i\hbar\mathbbm{1}m_2}{2(m_2+m_0)}$, which can then be rewritten in terms of $\{X_0, P_0, X_1, P_1,X_2,P_2\}$ with the help of the relations \eqref{eq:relativevariables}. The resulting expression reveals the type of operator in $S$ that corresponds to the action measured in $S'$.

The absence of a unitary operator implementing the desired transformation prevents us from deriving the active picture associated with the prescription \eqref{eq:prescription}. As a result, much of our intuition about quantum resources such as coherence and entanglement is lost. As a palliative solution to this issue, one may construct the state of each individual particle in the system relative to particle $0$. To this end, it suffices to perform the replacement $1 \to j$ in Eq.~\eqref{eq:T_cm,r}, and then trace over all partitions except $j$ to obtain $\rho_{r_j} = \Tr_{\forall k \neq j} (T_{\cm,r}^\dag\, \rho\, T_{\cm,r})$. Of course, the composition $\bigotimes_{j=1}^{N-1}\rho_{r_j}$ does not reproduce the $(N-1)$-partite relative state, which is generally entangled. In any case, for global pure states, the purity of $\rho_{r_j}$ will furnish a faithful diagnosis of whether the relative degrees of freedom associated with particle $j$ are entangled with the other partitions.

Given the significant boost observed in recent years in the literature on QRFs, we hope that our work highlights the need for better frameworks for transformations between physical reference frames and stimulates experiments that can guide such modeling.

\section*{Acknowledgments}
This study was financed in part by the Coordenação
de Aperfeiçoamento de Pessoal de Nível Superior - Brasil
(CAPES) - Finance Code 001. R.M.A. acknowledges financial support from the National Institute for Science and Technology of Quantum Information (CNPq, INCT-IQ 465469/2014-0) and the Brazilian funding agency CNPq under Grant No. 305957/2023-6.

\bibliography{references}
\end{document}